\documentclass[widereview]{vgtc}                  

\title{\systemnamelong: Visualizing the Broad Impact of Science and Science Funding
} 

\author{
Yifang Wang\textsuperscript{1,2,3,4,5},
Yifan Qian\textsuperscript{1,2,3,4},
Xiaoyu Qi\textsuperscript{6},
Yian Yin\textsuperscript{7},
Shengqi Dang\textsuperscript{6},
Ziqing Qian\textsuperscript{6},\\
Benjamin F. Jones\textsuperscript{1,2,3,4,8},
Nan Cao\textsuperscript{6*},
Dashun Wang\textsuperscript{1,2,3,4,9*}\\[1ex]
\textsuperscript{1}Center for Science of Science and Innovation, Northwestern University, Evanston, IL, USA\\
\textsuperscript{2}Northwestern Innovation Institute, Northwestern University, Evanston, IL, USA\\
\textsuperscript{3}Ryan Institute on Complexity, Northwestern University, Evanston, IL, USA\\
\textsuperscript{4}Kellogg School of Management, Northwestern University, Evanston, IL, USA\\
\textsuperscript{5}Department of Computer Science, Florida State University, Tallahassee, FL, USA\\
\textsuperscript{6}Intelligent Big Data Visualization Lab, Tongji University, Shanghai, China\\
\textsuperscript{7}Department of Information Science, Cornell University, Ithaca, NY, USA\\
\textsuperscript{8}National Bureau of Economic Research, Cambridge, MA, USA\\
\textsuperscript{9}McCormick School of Engineering, Northwestern University, Evanston, IL, USA\\
\textsuperscript{*}Correspondence to: \texttt{nan.cao@tongji.edu.cn},
\texttt{dashun.wang@northwestern.edu}
}

\abstract{
  Understanding the broad impact of science and science funding is critical to ensuring that science investments and policies align with societal needs. 
  Existing research links science funding to the output of scientific publications but largely leaves out the downstream uses of science and the myriad ways in which investing in science may impact human society. 
  As funders seek to allocate scarce funding resources across a complex research landscape, there is an urgent need for informative and transparent tools that allow for comprehensive assessments and visualization of the impact of funding. 
  Here we present \systemnamelong (\systemname), a visual analysis system for researchers, funders, policymakers, university leaders, and the broad public to analyze multidimensional impacts of funding and make informed decisions regarding research investments and opportunities. 
  The system is built on a massive data collection that connects 7M research grants to 140M scientific publications, 160M patents, 10.9M policy documents, 800K clinical trials, and 5.8M newsfeeds, with 1.8B citation linkages among these entities, systematically linking science funding to its downstream impacts. 
  As such, \systemnamelong is distinguished by its multifaceted impact analysis framework. 
  The system incorporates diverse impact metrics and predictive models that forecast future investment opportunities into an array of coordinated views, allowing for easy exploration of funding and its outcomes. 
  We evaluate the effectiveness and usability of the system using case studies and expert interviews. 
  Feedback suggests that our system not only fulfills the primary analysis needs of its target users, but the rich datasets of the complex science ecosystem and the proposed analysis framework also open new avenues for both visualization and the science of science research. 
}

\graphicspath{{figs/}{figures/}{pictures/}{images/}{./}} 

\usepackage{tabu}                      
\usepackage{booktabs}                  
\usepackage{lipsum}                    
\usepackage{mwe}                       

\usepackage{mathptmx}                  

\usepackage{balance}
\usepackage{enumitem}
\usepackage{setspace}
\usepackage{xspace}
\usepackage{microtype}                 
\PassOptionsToPackage{warn}{textcomp}  
\usepackage{textcomp}                  
\usepackage{mathptmx}                  
\usepackage{times}                     
\usepackage{cite}                      
\usepackage{tabu}                      
\usepackage{booktabs}                  
\usepackage{color}
\usepackage[dvipsnames,svgnames]{xcolor}
\usepackage[most]{tcolorbox}
\usepackage{graphicx}
\usepackage{multirow}
\usepackage{booktabs}
\usepackage{colortbl}
\usepackage{afterpage}
\usepackage{ragged2e}
\renewcommand{\footnotesize}{\raggedright\scriptsize}

\definecolor{variable-background}{RGB}{244, 244, 245}
\definecolor{variable-border}{RGB}{136, 136, 136}
\definecolor{variable-text}{RGB}{175, 175, 175}
\newtcbox{\variable}[1][]
  {on line, 
   colframe= variable-border, 
   colback = variable-background, 
   arc = 2pt, 
   boxsep = 0pt, 
   left = 2pt, right = 2pt, top = 2pt, bottom = 2pt, 
   boxrule = 1pt, 
  }

\newcommand{\systemname}{\textit{FtF}\xspace} 
\newcommand{\systemnamelong}{\textit{Funding the Frontier}\xspace} 
\newcommand{\queryview}{\textit{Query View}\xspace}
\newcommand{\grantview}{\textit{Grant View}\xspace}
\newcommand{\piview}{\textit{PI View}\xspace}
\newcommand{\mainview}{\textit{Impact Landscape View}\xspace}
\newcommand{\impactselectionview}{\textit{Impact Type View}\xspace}
\newcommand{\impactentityview}{\textit{Impact Entity View}\xspace}
\newcommand{\impactglyph}{\textit{ImpactGlyph}\xspace} 
 
\newcommand{\wordle}{\textit{TopicLens}\xspace}

\newcommand{\etal}{et al.\xspace}
\newcommand{\ie}{i.e.}
\newcommand{\eg}{e.g.}
\newcommand{\ea}{$E_{A}$\xspace} 
\newcommand{\eb}{$E_{B}$\xspace} 
\newcommand{\ec}{$E_{C}$\xspace} 
\newcommand{\ed}{$E_{D}$\xspace} 
\newcommand{\ee}{$E_{E}$\xspace} 
\newcommand{\newea}{$P_{A}$\xspace} 
\newcommand{\neweb}{$P_{B}$\xspace} 
\newcommand{\newec}{$P_{C}$\xspace} 
\newcommand{\newed}{$P_{D}$\xspace} 
\newcommand{\newee}{$P_{E}$\xspace} 
\newcommand{\newef}{$P_{F}$\xspace} 
\newcommand{\tasktypeone}{Project Summarization\xspace} 
\newcommand{\tasktypetwo}{Impact Evaluation\xspace} 
\newcommand{\tasktypethree}{Investment Recommendation\xspace} 
\newcommand{\tasktypeoneshort}{\textbf{S}} 
\newcommand{\tasktypetwoshort}{\textbf{E}} 
\newcommand{\tasktypethreeshort}{\textbf{R}}

\definecolor{wyftextcolor}{RGB}{237, 85, 106} 
\newcommand{\wyf}[1]{\textcolor{black}{#1}} 
\definecolor{dashuntextcolor}{RGB}{255, 153, 0}

\definecolor{todiscuss}{RGB}{237, 85, 106} 
\newcommand{\todiscuss}[1]{\textcolor{black}{#1}}

\usepackage{caption}
\makeatletter
\@ifundefined{captionfonts}
  {\newcommand{\captionfonts}{\normalsize\sffamily}}
  {\renewcommand{\captionfonts}{\normalsize\sffamily}}
\makeatother
\AtBeginDocument{%
  \captionsetup{font={small,sf}}
}

\usepackage{fancyhdr}
\begin{document}
\nocopyrightspace



\maketitle

\thispagestyle{fancy}   
\begin{spacing}{1} 
\newpage
\section{Introduction}
\label{sec:01_Introduction}  
Scientific progress is crucial to human well-being and prosperity. Advances in science have improved global health~\cite{williams2015scientific, wang2021pharmaceutical, thelwall2016citations}, driven innovation and economic growth~\cite{ahmadpoor2017dual, fleming2019government}, and informed policy decisions~\cite{national2012using, furnas2025partisan}. From mRNA vaccines to the Internet, scientific progress improves quality of life and standards of living, widely seen as a vital public good~\cite{bush2020science, azoulay2011incentives, yin2022public}. 
Essential drivers of scientific progress are funding and public support. 
Vannevar Bush highlighted this connection in his landmark 1945 report, \textit{``Science, The Endless Frontier,''} advocating for federal investment in scientific research as \textit{``the fund from which the practical applications of knowledge must be drawn''}~\cite{bush2020science}.  
This report led to the creation of the National Science Foundation (NSF) in 1950 and established the basis for modern science policy.

While there is now growing evidence of the strong relationship between science funding, basic science, and its societal impact at a high level~\cite{jones2021science, yin2022public}, a holistic and systematic approach to evaluating the broad outcomes of specific funding decisions remains scarce. Given the growing importance of science funding, the ability to evaluate the multidimensional impacts of this funding is critical to ensure that science policies and investments align with social needs and to maintain public trust in scientific endeavors~\cite{NSFBroaderImpacts}. These evaluations are of immense interest to a diverse array of stakeholders, including funding agencies seeking effective science investment strategies, national policymakers pursuing opportunities to enhance national competitiveness through science and technology policies, university leaders planning strategic initiatives, and individual scientists formulating impactful research programs. However, the analysis required for this work is complex and challenging due to both the multifaceted nature of funding outcomes and the extended time period that passes between the initiation of projects and their potential societal impact. 

Much research has emphasized the importance of science funding~\cite{price1963little, fortunato2018science, wang2021science}. Yet, the bulk of existing research concentrates on grants and their resulting scientific publications, which evaluate the impact of funding within science, while ignoring other facets of the impacts. Emerging efforts examine the impacts of scientific research more broadly, but they rarely trace these outcomes back to their funding sources~\cite{weller2015social, williams2015scientific, ahmadpoor2017dual, furnas2025partisan}. 
The visualization community has also long been interested in studying science funding, using, for example, grant data for topic mining and principal investigator (PI) network visualization~\cite{molnar2015using, dou2013hierarchicaltopics}, yet it, too, has largely overlooked the broad impact of funding. As a result, there is a notable lack of quantitative analyses and accessible analytical approaches that science decision-makers can utilize effectively, as well as a substantial gap in our understanding of the full landscape of the influence of funding on science and society.

\begin{figure} 
 \centering 
 \includegraphics[width=0.8\linewidth]{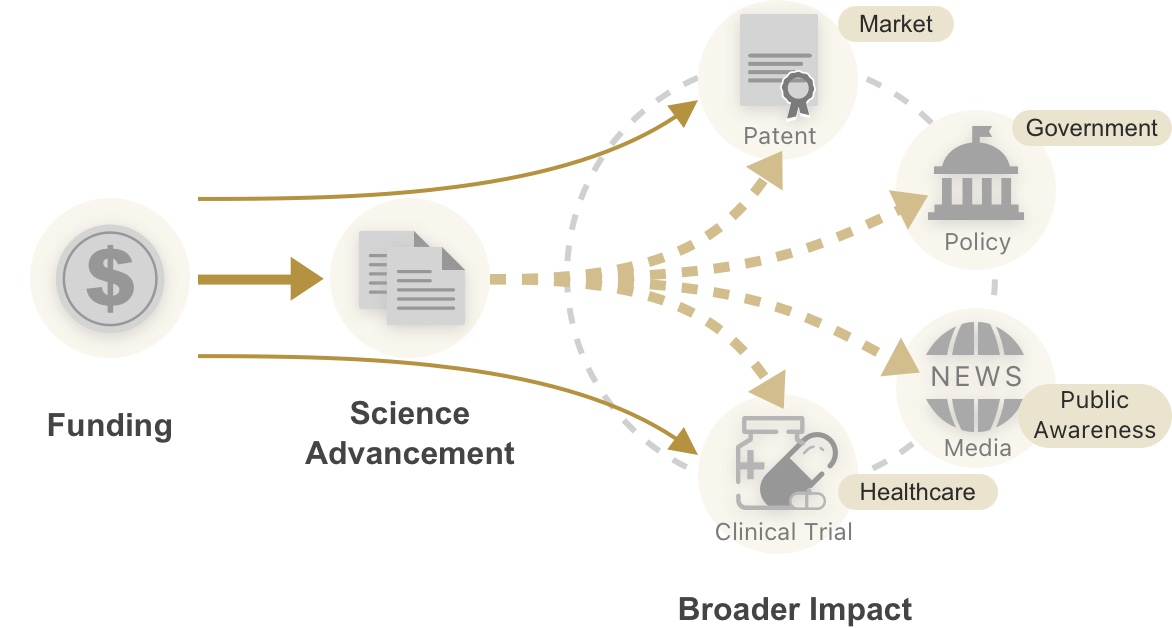}
 \caption{The science ecosystem, from the upstream funding to the science to the broader downstream impact. 
 \todiscuss{The direct outcomes of funding include scientific publications, patents, and clinical trials (solid dark yellow arrows), while the broader impacts emerge through the influence of funded papers on the broader society, such as public policy (dashed shallow yellow arrows).}
 } 
 \label{fig:background}
\end{figure}

In recent years, extensive new datasets have emerged that capture not only the inner workings of the scientific enterprise in detail and at scale, including funding~\cite{Dimensions} and scientific outputs, but also trace the numerous ways in which science interfaces with the broader human society, ranging from healthcare~\cite{AACT} to government policies~\cite{Overton} to marketplace applications~\cite{PatentsView} to public perceptions~\cite{Altmetric}, \wyf{representing science as an interconnected ecosystem (Fig.~\ref{fig:background})}. 
To truly unleash the potential of these datasets and to maximize the impact of science on society, however, we need new toolkits to analyze and understand the information that is now available to us. 
Here we present \systemnamelong (\systemname), a new visualization tool that combines the latest advances across two fields---the science of science (SciSci)~\cite{wang2021science, gates2023reproducible} and visual analytics~\cite{keim2008visual}. Indeed, SciSci methods enable us to link various data sources to create a holistic and systematic approach to assessing and predicting the broad impact of funding, and visual analytics solutions empower us to deliver accessible insights to policymakers and institutional leaders directly. 
Overall, the system builds on a massive data collection that connects 7M research grants to 140M scientific publications, 160M patents, 10.9M policy documents, 800K clinical trials, and 5.8M newsfeeds, with 1.8B citation linkages among these entities, representing to our knowledge the largest and most comprehensive data aggregation of science funding and its downstream impacts. 
By assessing the broad impact of science and science funding, the system allows for more targeted investments, an improved ability to relieve roadblocks in the scientific process, and, overall, a greater impact of science on society. 

There are several challenges involved in designing such a system. First, impact assessment requires rational definitions and accessible metrics for evaluating the multidimensional impacts of funding on society~\cite{NSFBroaderImpacts}. These measures need to capture the wide range of potential outcomes of scientific research, while being straightforward and comprehensible to decision-makers who often lack technical backgrounds. Second, the diversity and volume of the data create a need for especially innovative visualization solutions to allow for effective exploration. The data not only cover a wide range, from grant information to scientific outcomes and their downstream impacts, but also introduce a high level of complexity, as they are characterized by heterogeneous citation networks, multidimensional variables, hierarchical topics, and temporal dynamics. Third, a system that can offer meaningful investment recommendations and guide future investment strategies requires a predictive engine that can analyze the massive amount of historical data on funding and its outcomes and suggest high-impact opportunities.

We began to tackle these challenges by interviewing multiple experts, including both decision-makers and active researchers, to characterize the domain problem. We then defined a set of measures to quantify the impact of funding from multiple dimensions and developed a prediction model to identify grant topics and research investigators with high potential to produce substantial impacts in the future. Using these historical data and prediction results, we designed and developed \systemnamelong (\systemname), a comprehensive visual analysis system that presents, analyzes, and predicts the multi-faceted impacts of science and science funding for a wide range of decision-makers in science, including funders, policymakers, university leaders, and more. 
The contributions of this work are summarized as follows:
\begin{itemize}[leftmargin=10pt,topsep=1pt,itemsep=1px]
    \item We characterize the problem domain of visual analytics for funding impact and propose a list of impact measures. 
    \item We develop \systemname, a comprehensive visual analysis system with novel designs and effective interactions to explore multi-dimensional funding impact. 
    \todiscuss{The system is available through the website: \href{https://fundingthefrontier.com}{https://fundingthefrontier.com}.}
    \item We conduct an \todiscuss{exploratory} evaluation of the effectiveness and usability of the new system, including a quantitative study, two case studies, and expert interviews. 
\end{itemize}

\section{Related Work}
\label{sec:02_RelatedWork} 
Related literature spans three domains, including the science of science, funding data visualization, and network visualization. 

\subsection{The Science of Science} 
\label{sec:02_RelatedWork_FundingImpactAnalysis} 
The SciSci community has long been interested in studying science funding and its outcomes. 
The early literature mainly focuses on the funding data itself, without considering its outcomes, revealing for example the racial and gender bias in grant applications~\cite{forscher2019little, ginther2011race}. 
To the extent that funding outcomes are examined, the bulk of research has focused on the impact of funding within science itself, studying for instance how funding correlates with the amount and impact of papers that result~\cite{li2015big, wang2011funding, miao2023cooperation, azoulay2011incentives, vasan2021hidden}, or the impact of grants on subsequent career outcomes in science~\cite{wang2019early, bol2018matthew, azoulay2011incentives}. 

Separately, there is another line of inquiry that aims at quantifying the broad impacts of science. While these studies help us better understand how science interfaces with various aspects of human society, including marketplace applications~\cite{ahmadpoor2017dual}, public perception~\cite{Altmetric, weller2015social}, clinical trials~\cite{williams2015scientific, wang2021pharmaceutical, thelwall2016citations}, and policy-making~\cite{national2012using, furnas2025partisan, yin2022public} (Fig.~\ref{fig:background}), they do not trace the science back to its upstream funding, highlighting the empirical gap linking funding to its broad impacts.  

More recently, there have been some initial attempts at linking funding to its broad impacts, which substantially inform our work. 
For example, Yin \etal consider the broad impacts of public funding by examining three types of public uses of science (\ie, policy, news media, and patents)~\cite{yin2022public}. 
Other studies link government funding to resulting patents, highlighting the impact of funding on marketplace applications~\cite{li2017applied, fleming2019government}. 
While these papers rely on datasets that are more limited in scope, scale, and detail, compared with our datasets, they are similar to our work in that they address the relationship between funding and its broad impact. 
More importantly, while these studies are descriptive by nature, the policy relevance of the insights derived from these studies highlights the need for a visual analytical approach to understanding the broad impact of the funding, which is the focus of the present paper. 
Indeed, the key differentiation is that these existing studies are descriptive by nature, 
and do not allow decision-makers, such as program officers at funding agencies, to explore the data or draw insights. To the best of our knowledge, the visual analytics system we propose here is the first user-oriented, interactive system for science decision-makers to analyze and visualize the multidimensional impacts of funding.

\subsection{Scientific and Funding Data Visualization} 
\label{sec:02_RelatedWork_FundingDataVisualization} 
The visualization community has widely used and analyzed scientific data on papers, patents, and grants. 
Previous literature mainly falls into two categories: (1) technique-centered work and (2) insight-centered work. 
Using the complex data structure of scientific data, which includes network~\cite{dork2012pivotpaths, ye2022visatlas}, time-series~\cite{wang2021seek, yan2021turinggalaxy}, and multivariate~\cite{nobre2018juniper} datasets, 
technique-centered studies often use these data to propose general visualization techniques. 
Insight-based studies, by contrast, either use comprehensive visualization systems~\cite{li2019galex, dong2019vistory, guo2022sd, narechania2021vitality} or are based on datasets that are developed to distill insights about the development of the visualization and HCI community itself~\cite{chen2021vis30k, deng2022visimages, isenberg2016vispubdata, cao2023breaking}. 

More specifically, a number of studies have focused on science funding data. 
For example, DIA2~\cite{madhavan2014dia2} visualizes the NSF funding portfolio. 
Liu \etal~\cite{liu2006visual, liu2008interactive} proposed a 2.5D treemap to visualize funding distributions. 
Scholar Plot~\cite{kwonscholar} visualizes researchers' publication and funding data. 
The Public Innovations Explorer~\cite{schimmel2021public} visualizes funding from federal agencies and small businesses across the US to explore innovative geolocations. 
The EnArgus system~\cite{oppermann2021finding} provides semantic search functions for grant abstract data to focus on energy-related projects. 
And there are online platforms, including NIH RePORTER~\cite{NIHRePORTER, hesselberth2015nihexporter}, Altmetric~\cite{Altmetric}, PlumX Metrics~\cite{PlumXMetrics}, and Dimensions~\cite{Dimensions}, which serve primarily as search engines that show grants-related meta-information and their outputs.
Moreover, grant data have also been used to inform general visualization challenges, such as graph visualization using multidimensional grant metadata~\cite{ghani2012multinode, ghani2013visual} and topic modeling and evolution using grant abstract~\cite{dou2013hierarchicaltopics, dou2011paralleltopics}.
Related, there is a body of visualization work that analyzes funding outcomes in science~\cite{gilbert2011applying, lazar2013workshop, molnar2015using}. 

On one hand, these studies and systems highlight the importance of studying the impact of funding, but they have mainly focused on the impact within science, largely ignoring the impacts of funding on the broader human society. On the other hand, systems such as 
MedChemLens~\cite{shi2022medchemlens} and InnovationInsights~\cite{wang2023innovationinsights} explore the interaction between science and its broader impact, including patents and medicine, but they do not consider upstream funding.  
%
By contrast, our work characterizes the domain problem of analyzing funding impact and aggregates multiple data sources to develop a first-of-its-kind visualization system that evaluates the broad outcomes of funding portfolios, especially those that go beyond science.

\subsection{Network Visualization} 
\label{sec:02_RelatedWork_NetworkVisualization} 
Network visualization represents a broad area of study, covering a range of different network types, such as multi-variate~\cite{nobre2019state}, group~\cite{vehlow2015state}, dynamic~\cite{beck2017taxonomy}, and multi-layer networks~\cite{mcgee2019state}. 
Here we introduce studies that are most relevant to ours. 

The data in the scientific ecosystem can be depicted as a large-scale heterogeneous network comprising various node types (\eg, grants, papers, and researchers) and links (\eg, paper-authorship and grant-paper citations). 
For large-scale networks, strategies such as aggregation and graph sampling are commonly employed to manage visual complexity, often supplemented with interactions like filtering and zooming. Compound graphs~\cite{beck2017taxonomy} serve to group nodes into categories based on user-defined attributes or clustering methods. 
For example, OnionGraph~\cite{shi2020oniongraph} aggregates paper nodes into groups based on their citation range. \textit{FtF} also uses node aggregation to present the funding impact at the field level. 

When it comes to visualizing networks that include multiple node types, there are broadly four main approaches to visually differentiating node types within a network: embedding, superimposition, juxtaposition, and the use of visual node attributes~\cite{vehlow2015state}. 
To further convey hierarchy within a node type, contour overlays (\eg, circular treemaps~\cite{holten2006hierarchical}) have been shown to be intuitive and effective. 
However, these approaches often impose uniform data structures on each node type, which may neglect the diversity and the relationships across different node types. 
In \textit{FtF}, we adopt a hybrid approach that combines node-link diagrams with hierarchical structures and glyph techniques, which is tailored for assessing the multi-dimensional impacts of funding.

\section{System Overview}
\label{sec:03_Background} 
This section presents the analysis requirements and a system overview. 

\subsection{Analysis Tasks} 
\label{sec:03_Background_AnalysisTask} 
Conversations with diverse stakeholders in the past few years underscored the urgent need to evaluate the impact of funding to enhance decision-making processes, such as funders' investment strategies. These stakeholders included directors and program officers in private and public funding agencies, leaders at top universities, policymakers, and SciSci researchers. We identified key challenges these decision-makers face, including the absence of comprehensive data on funded programs and their wide-ranging impacts, the lack of systematic approaches to measuring the impact of funding, and the scarcity of user-friendly tools for those who lack specialized data analysis skills. These discussions highlighted the need for a visual analytics system capable of providing an intuitive, systematic, and efficient approach to evaluating the impacts of funding in order to enhance decision-making.

We began to address this challenge over the past few years, engaging with experts from various sectors. 
Our experts included a U.S. federal funding agency program officer with over eight years of experience (\ea), a director-level executive in a top private investment firm that specializes in investing in science (\eb), 
as well as experts in the SciSci field (\ec-\ee). \ec is an established SciSci professor and \ed is a professor who works on statistical analysis of funding data. \ee is a researcher focusing on science and innovation. 
\todiscuss{We also gathered requirements and feedback from data analysis teams at the National Science Foundation (NSF) and the National Institutes of Health (NIH).} 
Based on conversations with decision-makers, discussions with domain experts, and a comprehensive literature review, we leverage the expert-focused design study methodology~\cite{sedlmair2012design} to identify the six analysis tasks listed below.

\textbf{Tasks for \tasktypeone (\tasktypeoneshort)}. 
Profiles of funded projects offer a strategic entry point for analysis, enabling users to examine the investments and adjust strategies accordingly. Therefore, the system should provide both overview and detailed information about the funded projects and support a flexible information exploration approach.
\begin{enumerate}[topsep=1pt,itemsep=2px,partopsep=2pt,parsep=2pt]
    \item[{\bf \tasktypeoneshort1:}] {\bf Dynamic Data Selection.} The system should enable users to dynamically filter grants by various attributes, such as funding agencies and grant years, to tailor their analysis. 
    \item[{\bf \tasktypeoneshort2:}] {\bf Funding Overview Illustration.} The system should provide a summary of funded fields to assist analysts in selecting fields of interest for detailed analysis. 
    \item[{\bf \tasktypeoneshort3:}] {\bf Researcher and Institution Characterization.} The system should also characterize (1) the funded researchers (\ie, principal investigator (PI)), in terms of demographics and SciSci measures of research ability, and (2) recipient research institutions. This information supports diversity, equity, and inclusion (DEI) by highlighting the funding distribution among researchers and institutions and identifying PIs and institutions with potential for future investment. 
\end{enumerate}

\textbf{Tasks for \tasktypetwo (\tasktypetwoshort)}. 
Another key analysis requirement is the ability to assess the impact of funded projects. As funding agencies have a wide range of missions and investment strategies, the criteria for ``success'' in different funding portfolios can vary significantly.
The analysis should, therefore, present the impact of funding from a multidimensional perspective and provide context to aid users in understanding these outcomes.
\begin{enumerate}[topsep=1pt,itemsep=2px,partopsep=2pt,parsep=2pt]
    \item[{\bf \tasktypetwoshort1:}] {\bf Analysis of funding impacts.} The system should offer a comprehensive set of easy-to-understand metrics to help analysts evaluate and select the type of impact they value most---ranging from direct scientific impacts to broader societal benefits. Intuitive visualizations and interactions should be integrated to present more detail regarding the impact, which is based on the citation relationships between grants and their outcomes (\eg, papers). 
    \item[{\bf \tasktypetwoshort2:}] {\bf Contextualization of funding impacts.} To deepen users' understanding of funding outcomes and enrich the insights from the analysis, the system should go beyond citation relationships to incorporate relevant contextual data, including detailed information about the attributes of specific outcomes, such as the drugs that result from clinical trials or the assignees of the patents.
\end{enumerate}

\textbf{Tasks for \tasktypethree (\tasktypethreeshort)}. 
The ultimate aim of these evaluations of the impact of funding is to identify promising directions for future investments. Predictive analytics can play a pivotal role in guiding these forward-looking decisions.
\begin{enumerate}[topsep=1pt,itemsep=2px,partopsep=2pt,parsep=2pt]
    \item[{\bf \tasktypethreeshort1:}] {\bf Identifying Investment Opportunities.} Our system should support a forward-looking capability, allowing users to uncover potential investment opportunities based on their desired impact outcomes for the future. This involves pinpointing emerging research topics and, more importantly, identifying the right people (PIs) for collaboration or funding. 
\end{enumerate}

\begin{figure} 
 \centering 
 \includegraphics[width=0.8\linewidth]{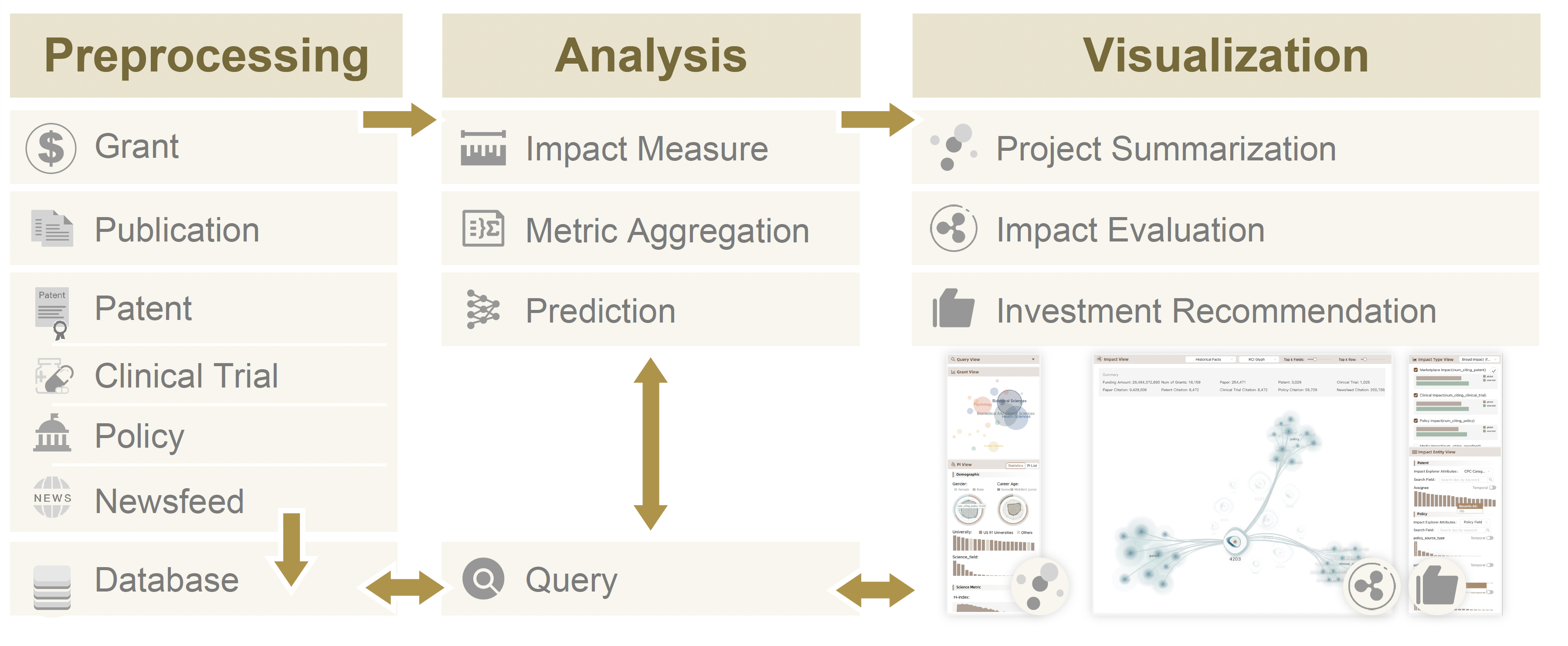}
 \caption{\systemname system overview. The system consists of a preprocessing module, an analysis module, and a visualization module.} 
 \label{fig:system-overview}
\end{figure}

\subsection{System Overview} 
\label{sec:03_Background_SystemOverview} 
Guided by these analysis tasks, we designed \systemname, a web-based application that consists of three modules \wyf{(Fig.~\ref{fig:system-overview})}: (1) the preprocessing module, (2) the analysis module, and (3) the visualization module. 
The preprocessing module cleans multiple data sources, stores them in the database, and supports dynamic data queries (\textbf{\tasktypeoneshort1}). 
The analysis module provides a list of metrics (\textbf{\tasktypeoneshort3}, \textbf{\tasktypetwoshort1}) and prediction models to indicate grant topics, principal investigators (PI), and research institutions for future investment (\textbf{\tasktypethreeshort1}). These two modules form the backend of the system, which is implemented by Google BigQuery, Python, and Flask. 
The visualization module uses coordinated views with intuitive interactions to present both historical data on grant outcomes (\textbf{\tasktypeoneshort2}, \textbf{\tasktypeoneshort3}, \textbf{\tasktypetwoshort1}, \textbf{\tasktypetwoshort2}) and prediction results (\textbf{\tasktypethreeshort1}) to guide future grant-making, implemented by React.js, Redux.js, D3.js, and TypeScript. 
We introduce each module with more details in the following sections. 
\section{Data Analysis}
\label{sec:04_DataAnalysis} 
The analysis module is responsible for the calculations that serve as the foundation for the visual analytics systems. First, this module assesses and aggregates the multidimensional impacts of science and funding at different levels, from individual papers and grants to broader categories at the PI and field levels. Second, the module predicts future research investment opportunities by estimating the potential impact of recent grants. We present more details regarding the system's data sources, preprocessing methods, funding impact assessments, and future investment recommendations below.

\subsection{Data Sources and Preprocessing}
\label{sec:04_DataAnalysis_DataPreprocessing} 
This analysis of the broad impact of science funding requires the integration of the following four datasets containing information about the science ecosystem: 
\begin{itemize}[leftmargin=10pt,topsep=2pt,itemsep=1px]
    \item \textbf{Dimensions}~\cite{Dimensions}. This dataset comprises data on global science funding (7M grants), scientific publications (140M), patents (160M), clinical trial records (800K), and their citation linkages (1.7B). It also contains information about researchers (51.7M), including their names and affiliations and their connections to grants (9M, mostly PIs) and paper-author pairs (461M). 
    \item \textbf{Overton}~\cite{Overton}. This dataset includes global policy documents (10.9M) and citation linkages between policies and papers (13M). 
    \item \textbf{Altmetric}~\cite{Altmetric}. This dataset provides insights into public engagement with science through newsfeeds (5.8M) and citations between news sources and papers (6M). 
    \item \todiscuss{\textbf{SciSciNet}~\cite{lin2023sciscinet}. 
    This dataset includes a large-scale open data lake for the science of science research, covering over 134M scientific publications and millions of external linkages to funding and public uses. 
    Specifically, we use a set of SciSci metrics (\eg, h-index and productivity) at the paper and researcher level, which is a great addition to the Dimensions dataset.}
\end{itemize}

We preprocessed the data sources to limit our analysis to all types of documents published between 2000 and 2021 to ensure robust data coverage (see Supplementary Notes 1 and 2). 

\begin{figure*} [!htb]
 \centering 
 \includegraphics[width=1\linewidth]{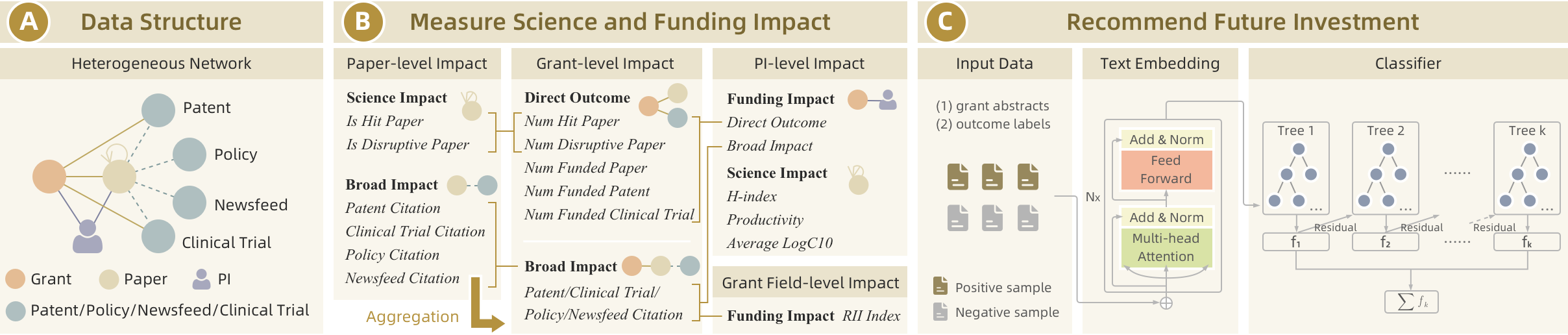}
 \caption{The data analysis procedure. 
 (A) We preprocess the data into a heterogeneous network with multidimensional attributes. 
 (B) We then define and calculate a list of metrics to measure the impact of science and funding at multiple levels. 
 (C) Finally, we use a prediction model to recommend the grant topics and PIs \wyf{that represent the best opportunities to achieve future impact in a specific impact topic}. 
 } 
 \label{fig:framework}
\end{figure*}

\subsection{Measure the Funding Impact} 
\label{sec:04_DataAnalysis_ScientificMeasure}
The impact of science and science funding is multidimensional and extends through a range of levels, from papers and grants at the lowest level to grant fields and funding agencies at the highest level. In collaboration with experts in the SciSci field, we developed a set of metrics to assess the diverse funding outcomes. These metrics are multidimensional, measuring not only the direct scientific impact of the funding based on publications and citations but also measuring a wide range of broader impacts, encompassing the marketplace, clinical trials, policy, and media. These impacts can be categorized into the direct outcomes of the funding and its broader impacts, which are distinguished by different citation relationships: 

\textbf{Direct Outcomes.} The direct impact of funding can primarily be seen in scientific publications, patents, and clinical trials that cite grants directly. These outcomes are captured by the number of citations in papers, patents, and clinical trials to a grant. For a given grant $\mathcal{G}$, its direct impact can be measured by: 
\begin{itemize}[leftmargin=10pt,topsep=1pt,itemsep=0px]
    \item \underline{\textit{Number of Funded Papers}}: the total number of published papers that acknowledge the funding $\mathcal{G}$. 
    \item \underline{\textit{Number of Funded Hit Papers}}: the number of hit papers that acknowledge $\mathcal{G}$. A hit paper is a paper that has a citation count, $C_p$, that is above a particular percentile threshold $T_y^f$ (\eg, top 5\%) in a specific field $f$ and year $y$. 
    \item \underline{\textit{Number of Funded Disruptive Papers}}: the number of funded papers within the top 5\% based on their disruption index, a measure derived from citation networks~\cite{wu2019large}. 
    The disruption index ($D$) for a paper $p$ quantifies its divergence from the existing literature, potentially indicating the opening of new research areas. It is calculated as: $D_p = \frac{n_i - n_j}{n_i + n_j + n_k}$, where $n_i$ is the number of future papers that cite the focal paper $p$ but none of its references, $n_j$ is the number of future papers that cite the paper $p$ and one or more of its references, and $n_k$ is the number of references in $p$ that are not cited by papers citing $p$. 
    \item \underline{\textit{Number of Funded Patents}}: the total number of patents that acknowledge $\mathcal{G}$. 
    \item \underline{\textit{Number of Funded Clinical Trials}}: the number of clinical trials that acknowledge $\mathcal{G}$. 
\end{itemize}

\textbf{Broader Impact.} Recent studies suggest that the true impact of funding goes far beyond publications, affecting industry and society more broadly~\cite{yin2022public}. To capture these broad outcomes from funding, we examine the citation linkages between funded papers and various downstream applications. 
By doing so, we not only expand the pools of patents and clinical trials stemming indirectly from funding~\cite{li2017applied}, but also track the broad impact of science on other public sectors such as policymaking and media news. 
Given a funded paper $\mathcal{P}$, we define the following metrics to estimate its broader impact: 
\begin{itemize}[leftmargin=10pt,topsep=1pt,itemsep=0px]
    \item \underline{\textit{Patent Citation}}: the total number of patents citing $\mathcal{P}$. This metric represents the marketplace impact of a paper. 
    \item \underline{\textit{Clinical Trial Citation}}: the total number of clinical trials citing $\mathcal{P}$, measuring the clinical impact. 
    \item \underline{\textit{Policy Citation}}: the total number of policy documents citing $\mathcal{P}$, measuring the policy impact. 
    \item \underline{\textit{Newsfeed Citation}}: the total number of news articles citing $\mathcal{P}$, measuring the media impact. 
\end{itemize}
Using the grant-to-paper linkages, we aggregate the impact metrics for all individual papers that cited a grant. 
We define the broader impact of a grant by summing up all such downstream citation counts received by papers funded by the grant. 

\textbf{Aggregation.} Based on grant-level impact, we further aggregate and calculate the funding impact at the PI level and grant field level by summing the direct and broader impact measures of individual grants. Additionally, we calculate the following three metrics for a PI to measure the PI's scientific impact: 
\begin{itemize}[leftmargin=10pt,topsep=1pt,itemsep=0px]
    \item \underline{\textit{H-index}}: the PI's h-index. The index of a researcher is h if she has h papers with at least h citations and all her remaining papers have fewer than h citations. 
    \item \underline{\textit{Productivity}}: the PI's total number of publications. 
    \item \underline{\textit{Average LogC10}}: the log value of the average number of citations within 10 years of publication for all of the PI's papers~\cite{sinatra2016quantifying}. 
\end{itemize} 

We also calculate a relative impact index for a grant field or a funding agency \wyf{$\mathcal{X}$}. We introduce the \wyf{Relative Impact Index (RII)} to benchmark their performance across various impact types, compared to a global baseline. 
For a given impact type $i$ and a grant field or a funding agency $\mathcal{X}$, \wyf{RII measures the fraction of grants in $\mathcal{X}$ with an impact in an impact type $i$, normalized by the same fraction obtained on all grants for that impact type}: 
%
\begin{equation}
    RII_{i, \mathcal{X}} = \frac{\text{num\_grant} \ in\ \mathcal{X}\ with\ impact\ type\ i / \text{num\_grant}\ in\ \mathcal{X}}{Total\ \text{num\_grant}\ with\ impact\ type\ i / Total\ \text{num\_grant}} 
\end{equation}

\subsection{Scientific Investment Recommendation} 
\label{sec:04_DataAnalysis_Prediction} 
One powerful component of \systemname is its capability to inform funding agencies about future investment opportunities. This aspect of the system allows funders who seek a particular outcome to identify optimal topics and principal investigators (PIs) for future grants. As a substantial amount of time must pass after a grant is given to observe evidence of the impact of the funding, we employ a machine learning model that forecasts the future impact of recent grants across various impact domains and topics, thereby \wyf{allowing real-time insights into a recent} grant's likely outcomes and allowing for the identification of emerging high-impact grant topics and PIs within a chosen impact domain. 

In particular, we introduce a prediction model that utilizes grant abstracts as input to predict the intrinsic potential for various types of impacts, encompassing both direct outcomes and broader impacts, as detailed in Section~\ref{sec:04_DataAnalysis_ScientificMeasure}. For instance, focusing on patents as direct outcomes, our prediction task is to assess the potential of a grant to directly result in at least one patent in different patent categories, with each patent topic requiring a specific machine learning model. To this end, we leverage SciBERT~\cite{Beltagy2019SciBERT}, which is a large language model based on the BERT architecture and has been pre-trained on a large corpus of scientific text data, enabling it to understand text within the scientific domain. SciBERT transforms words in a grant abstract into tokens, each of which is mapped to a 768-dimensional vector. These vectors are then averaged to generate a single vector representing the entire abstract. This resultant vector can be employed as input for machine learning classifiers. One such classifier is XGBoost~\cite{chen2016xgboost}, an efficient and scalable implementation of gradient boosting algorithms. XGBoost is known for its speed and performance in handling large-scale datasets and has been widely used in various machine-learning tasks due to its high accuracy and interpretability. Here, we utilize XGBoost as the classifier, which takes the SciBERT embeddings as the input and determines whether a grant generated the impact regarding the specific prediction task. Additionally, we employ patents as direct outcomes to compare the performance of XGBoost with other common classifiers (e.g., random forest). Our findings indicate that XGBoost demonstrates comparable performance with other classifiers.

\textbf{Implementation Details.} A number of impact prediction models have been trained in our system for different impact factors. To facilitate the understanding, we take the ``direct patent outcome'' prediction as an example. Other models are trained by following a similar procedure.

In particular, we employ the pre-trained SciBERT model provided by the Allen Institute for AI~\cite{SciBERTImplementation}. This model takes a text input with no more than 512 words. It is an uncased model that does not distinguish between uppercase and lowercase letters during training or inference. We use it to convert the grant text into latent vectors. In the next step, we utilize the official Python implementation of XGBoost~\cite{XGBoostImplementation} in our system as the prediction model. The model has been trained based on a binary classification task with the goal of estimating the probability of a given grant having patent outcomes in topic $\mathcal{A}$ in the future.

To train the model, we collected all grants either from a funding agency or from a funding field. Based on these data, we calculate the averaged time cost (denoted as $Y$) for filing a patent with the funding support from the grant. After that, $Y$ is rounded as an integer to facilitate calculation. All the grants from $2000$ to $2021-Y$ have been used as the training samples for our prediction model. Here, $2021-Y$ gives a reasonable time span for filing the patent, mitigating potential biases from using grants with insufficient time windows. After that, we prepared both the positive and negative samples. In particular, given a topic $\mathcal{A}$, we collect all the related grants based on which one or more patents have been filed from $2000$ to $2021-Y$ as the positive samples, whereas an equal number of grants without any patent outcomes are randomly sampled as negative samples. 80\% of the positive and negative samples were used as training samples, leaving the rest 20\% for testing.


\begin{figure*} [!htb]
 \centering 
 \includegraphics[width=1\linewidth]{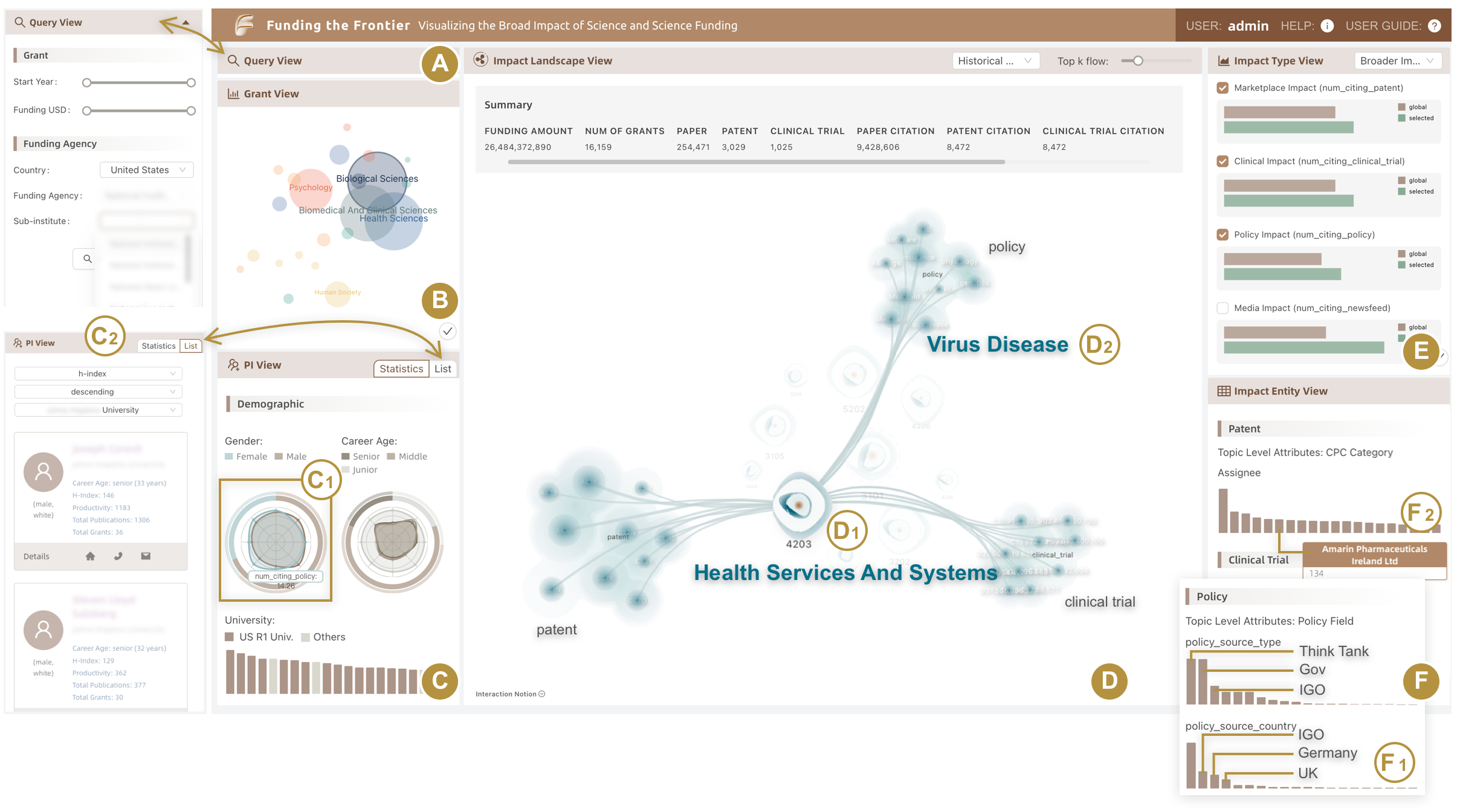}
 \caption{The \systemname system UI. The \queryview  (A) is for data filtering. The \grantview (B) and \piview (C) present grant project summaries. The \mainview (D) allows for a detailed exploration of the multidimensional impact of funding. The \impactselectionview (E) enables users to select the type of impact they want to explore, and the \impactentityview (F) displays contextual information about different types of impacts.}  
 \label{fig:system-ui}
\end{figure*}

\section{Visualization Design}
\label{sec:05_VisualDesign} 
In this section, we introduce the detailed visualization design of our system. We first briefly introduce the user interface as an overview and then describe the specific details of each visualization view. 

\subsection{User Interface}
\label{sec:05_VisualDesign_UserInterface} 
Our system integrates six coordinated views to support the analysis tasks outlined in Section~\ref{sec:03_Background_AnalysisTask} \wyf{(Fig.~\ref{fig:system-ui})}. Given the large amount of grant data, our system always lets the users start by filtering the data via the \queryview (\textbf{\tasktypeoneshort1}, Fig.~\ref{fig:system-ui}(A)), in which they can choose a funding agency or select a time range. The selected grants are shown in the \grantview (\textbf{\tasktypeoneshort2}, Fig.~\ref{fig:system-ui}(B)), in which the grants are clustered based on different research fields as bubbles with the bubble size indicating the funding amount. Once a field is selected, the details such as demographics and profiles of the corresponding PIs are visualized in the \piview (\textbf{\tasktypeoneshort3}, Fig.~\ref{fig:system-ui}(C)) for users' reference. At the same time, the corresponding grants, as well as their outcomes (i.e., policies, patents, clinical trials, and news reports that cite these grants, indicating the direct or broader impact of these grants), are visualized in the \mainview (\textbf{\tasktypetwoshort1}, Fig.~\ref{fig:system-ui}(D)) in the form of a multi-partite graph. In particular, all the grants are aggregated by research topics and shown in the center as glyphs. Their outcomes are also aggregated by topics and visualized as different clusters of graph nodes surrounding and connecting to the center glyphs. The sizes of the clusters directly estimate the grants' impact in terms of the corresponding outcomes. Users can hover on the grant or impact topic nodes and use a word cloud to learn fine-grained text information about this topic \wyf{(Fig.~\ref{fig:case1}(A, B))}. Users can switch between direct and broader impacts from the \impactselectionview (\textbf{\tasktypetwoshort1}, Fig.~\ref{fig:system-ui}(E)). The outcome clusters in the \mainview will be changed accordingly. 
The system also incorporates a context view showing the outcome distribution over other dimensions, such as policy source organizations (\textbf{\tasktypetwoshort2}, Fig.~\ref{fig:system-ui}(F)). 

Despite the above design that supports impact exploration, our system also predicts the potential impacts of each recent grant. The topics (i.e., the nodes in the impact graph) of the grants that may have a higher impact in the future are highlighted by purple rings (\textbf{\tasktypethreeshort1}, Fig.~\ref{fig:case1}(C)). The corresponding PIs are also listed in Fig.~\ref{fig:system-ui}(C2) for users (funding agencies) to make a future investment.

\subsection{Impact Landscape View}
\label{sec:05_VisualDesign_MainView}  
We designed the \mainview to illustrate the multidimensional impact of grants, using citation links between the grants and their outcomes in the form of a multipartite impact graph (\textbf{\tasktypetwoshort1}, Fig.~\ref{fig:system-ui}(D)). A grant may have four different types of outcomes \wyf{(\ie, direct or broader impact)}, including policies, patents, clinical trials, and news reports that acknowledge or mention the grant \wyf{or cite the funded papers of that grant}. These outcomes reveal the grant's impact from different aspects. To ensure scalability, both the grants and their outcomes are aggregated by their topic labels provided in our data. 
\wyf{As a result, in the impact graph, each node either indicates a collection of grants under the same research topic (denoted as the \underline{\textit{\textbf{grant topic node}}})} or a collection of outcomes indicating the same type of impact and sharing the same topic (denoted as the \underline{\textit{\textbf{impact node}}}). The node size represents the number of grants or outcomes in the cluster.

\paragraph{\bf Layout.}
To layout the graph, we introduce a hybrid layout method. 
In particular, we first use a force-directed layout method to layout the entity-type level graph (\eg, grant, paper, and patent). 
Second, within each impact type (\ie, paper, patent, clinical trial, and newsfeed), we employ a bubble treemap layout~\cite{gortler2017bubble} to pack the same type of impact nodes together as a cluster with a hierarchical topic structure. 
The centers of these bubble treemaps are then positioned at the locations of their corresponding entity-type nodes within the entity-type level graph.
Third, the cluster of grant topic nodes is initially placed in the middle of the view, surrounded by clusters of different impact nodes that are placed at the edge of the view. 
Another force-directed layout is employed to fine-tune the positions of the grant topic nodes to better reveal their relationships with a visually coherent and balanced layout. 
Specifically, the following forces are introduced to fine-tune the initial layout of the grant topic nodes:

\begin{itemize}[leftmargin=10pt,topsep=1pt,itemsep=0px]
\item \underline{\textit{Impact Force ($F_{impact}$):}} an attraction force between a grant topic node in the middle and the impact nodes in the surrounding clusters. It is formally defined as: 


\begin{equation}
    F_{\mathrm{impact}}
    = \sum_{i=1}^{n}
    \begin{cases}
    \bigl(RII_i - 1\bigr)\,\cdot d_i \,\cdot \hat u_i, & RII_i \ge 1,\\[4pt]
    \bigl(RII_i - 1\bigr)\,\cdot (1/d_i)\,\cdot \hat u_i, & RII_i < 1,
    \end{cases}
\end{equation}
where $F_{\text{impact}}$ is the impact force, $RII_{i}$ is the RII value for the i-th impact factor, 
$d_i = \lVert X_{\mathrm{impact}_i} - X_{\mathrm{grant}} \rVert$ is the Euclidean distance between the $i$-th impact type node and the grant node, 
$\hat u_i = \bigl(X_{\mathrm{impact}_i} - X_{\mathrm{grant}}\bigr) / d_i$ is the unit direction vector, 
$X_{\text{impact}_i}$ is the position vector of the i-th impact node, 
and $X_{\text{grant}}$ is the position vector of the grant glyph. 

\item \underline{\textit{Containment Force ($F_{contain}$):}} the force to constrain the positions of the grant topic nodes within the boundary of its containing cluster. It pulls a node towards the center if it strays beyond a predefined maximum distance. 
\begin{equation}
    F_{\text{contain}} = - \cdot \max(0, d - d_{\text{max}}) \cdot \hat{r},
\end{equation}
where $d$ is the distance from the node to the center, $d_{\text{max}}$ is the maximum allowed distance, and $\hat{r}$ is the unit vector pointing from the center to the node. 

\item \underline{\textit{Collision Force ($F_{collide}$):}}
a force that prevents nodes from overlapping with a padding parameter $p$. It is proportional to the amount of overlap between nodes: 

\begin{equation}
    F_{\text{collide}} = \cdot \max(0, r_1 + r_2 + p - d) \cdot \hat{r},
\end{equation}
where $F_{\text{collide}}$ is the collision force, $r_{1}$ and $r_{2}$ are the radii of the two nodes, $d$ is the distance between the centers of the two nodes, $p$ is a constant padding term, and $\hat{r}$ is the unit vector pointing from the center of one node to the center of the other node. 
\end{itemize} 

The three forces are balanced by three parameters to adjust the primary effect of the final layout: 
\begin{equation}
    \vec{F}_{\text{total}} = \alpha \vec{F}_{\text{impact}} + \beta \vec{F}_{\text{contain}} + \gamma \vec{F}_{\text{collide}},
\end{equation}
where $\vec{F}_{\text{total}}$ is the total force acting on the node, $\alpha$, $\beta$, and $\gamma$ are parameters that control the strength of the impact force, containment force, and collision force, respectively, and
$\vec{F}_{\text{impact}}$, $\vec{F}_{\text{contain}}$, and $\vec{F}_{\text{collide}}$ are the impact force, containment force, and collision force, respectively. 

We simulate the force system with alpha cooling and velocity decay, allowing it to gradually stabilize into a visually coherent and well-balanced layout. 
Finally, the citation linkages between the grant topic nodes and the impact nodes are bundled by impact types to reduce visual clutter and facilitate reading.

\begin{figure} 
 \centering 
 \includegraphics[width=0.8\linewidth]{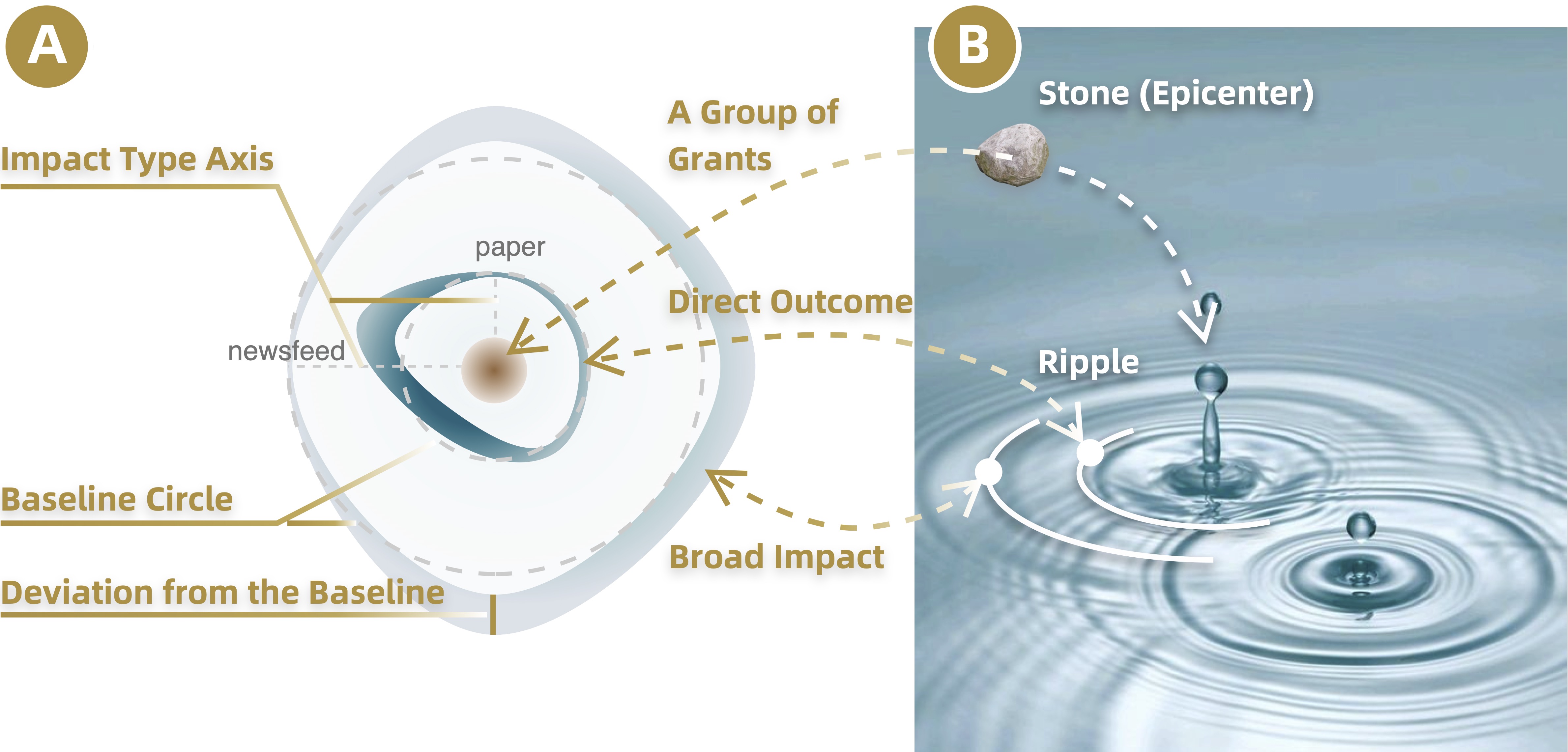}
 \caption{The visual design of the \impactglyph is inspired by the metaphor of a ripple. (A) In the historical mode, the glyph consists of a central grant circle surrounded by two concentric impact ripples. (B) An illustrative ripple metaphor, with a stone as the ``epicenter.''} 
 \label{fig:visual-design-glyph}
\end{figure}

\paragraph{\bf \impactglyph.} 
To summarize the multidimensional impact of each collection of grants represented by the grant topic node and facilitate the impact comparisons across different grant topics, the \impactglyph is introduced.  The design is inspired by a visual metaphor of the ripple caused by throwing a stone into the water, as shown in Fig.~\ref{fig:visual-design-glyph}.

In particular, the glyph consists of a center node (i.e., the collection of grants in the same topic) and two levels of outer rings respectively, indicating the corresponding direct and broader impacts of these grants. Two grey circular dashed lines set the global baseline for the RII index for the direct and broad impacts, respectively. Two radar charts symbolize the direct and broader impacts as concentric ripples from multiple dimensions, including patents, clinical trials, policies, and news mentions. In our design, we use a belt with thinness to replace the polyline that indicates multidimensional values in the traditional radar chart. The thickness of the belt encodes the grants' impact deviations across different impact dimensions from the baseline, which magnifies the difference. The color of the belt transitioning from darker to lighter green illustrates the grants' spillover effect from immediate outcomes to wider societal impacts. 
Altogether, this visualization enables users to swiftly grasp the impact of a group of grants across various dimensions relative to the global baseline using the shape and color of the ripple. 

In our system, the above glyph design has two modes: (1) the historical mode and (2) the prediction mode. In the historical mode (Fig.~\ref{fig:visual-design-glyph}), the historical funding impact is compared with global baselines using the RII index. The size of the center node reflects the number of historical grants within the timeframe selected by the user, serving as the ``epicenter'' of the ripple. 
In the prediction mode (Fig.~\ref{fig:case1}(C)), after specifying an impact topic of interest, the number of recent grants with high-prediction scores (above a threshold) will be highlighted as a purple ring, wrapping the previous historical center node, with the size encoding the number of high-prediction score grants.
The two impact ripples are hidden for simplicity. 

\paragraph{\bf \wordle.} Despite the above views, a \wordle (Fig.~\ref{fig:case1}(A, B)) is designed to illustrate the detailed topic keywords in the form of a wordle when the mouse hovers over a grant topic node in the impact graph. Word size encodes the overall keyword frequency with an overlaid sparkline showing its temporal trend, which provides insights into evolving grant or impact topic foci over time.

\subsection{Contextual Views}
\label{sec:05_VisualDesign_SideView} 
Despite the above primary views, \systemname provides a list of side views with contextual information, enabling users to understand the funding impacts and make informed decisions for future investments.  

\textbf{\queryview.} In this view, users can filter grants based on criteria such as grant year, funding amount, and agency, enabling the flexible search for grants of interest (\textbf{\tasktypeoneshort1}, Fig.~\ref{fig:system-ui}(A)). 

\textbf{\grantview.} This view summarizes the funded projects for the selected grants in terms of grant fields or sub-institutions (\textbf{\tasktypeoneshort2}, Fig.~\ref{fig:system-ui}(B)). The size of the bubble encodes the total funding amount within each grant field or sub-institution. 
The bubble's position indicates the similarity among grant groups (\ie, the semantic similarity of grant fields and funded grant fields of sub-institutions), as determined by dimension reduction techniques (\ie, t-SNE). 

\textbf{\piview.} This view shows the statistics of funded PIs, including demographics, universities, fields of study, and impact metrics (\textbf{\tasktypeoneshort3}, Fig.~\ref{fig:system-ui}(C)). In addition, we also provide a query and ranking using universities and impact metrics to enable locating detailed information in the PI list. 

\textbf{\impactselectionview.} This view provides a list of impact types for users to select, which will be further explored in the \mainview (\textbf{\tasktypetwoshort1}, Fig.~\ref{fig:system-ui}(E)). For each impact type, a bar chart shows the average number of impact documents \wyf{produced by} a grant against a global baseline. Users can switch between direct and broader impacts via a dropdown menu. 

\textbf{\impactentityview.} In addition to topics, this view provides context through bar charts summarizing the distribution of impact documents across various dimensions (\textbf{\tasktypetwoshort2}, Fig.~\ref{fig:system-ui}(F)), such as patent assignees, policy sources, clinical trial phases, and newsfeed origins, aiding in a deeper understanding of funding impacts.


\section{Evaluation}
\label{sec:06_Evaluation} 
We evaluated \systemname through a quantitative study of the prediction model, two case studies, and a series of interviews with experts.

\begin{figure}[tb]
  \centering
  \begin{subfigure}[b]{0.5\columnwidth}
  	\centering
  	\includegraphics[width=0.8\textwidth]{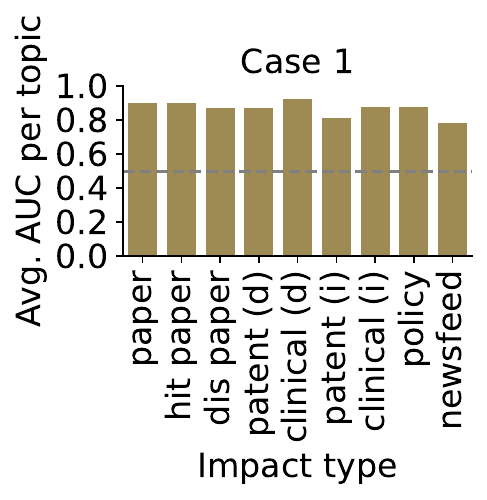}
  	\label{fig:case1_auc}
  \end{subfigure}%
  \hfill%
  \begin{subfigure}[b]{0.5\columnwidth}
  	\centering
  	\includegraphics[width=0.8\textwidth]{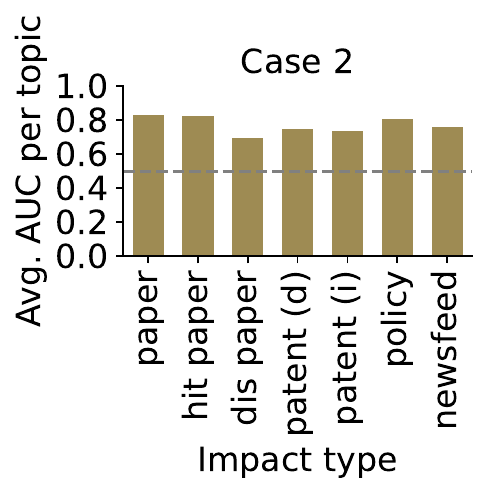}
  	\label{fig:case2_auc}
  \end{subfigure}%
  \subfigsCaption{The prediction performance (AUC) on the test set for two case studies. The average AUC is reported for each type of impact, based on the performance across topics for that type of impact. The dashed lines at 0.5 indicate the baseline performance of random guesses.
  Overall, the results demonstrate good performance for our prediction tasks. (d) represents direct impact and (i) represents indirect impact.
  }
  \label{fig:case1_case2_auc}
\end{figure}

\subsection{Quantitative Evaluation of the Prediction Model} 
We evaluated our model's performance using the Area Under the Curve (AUC) metric and assessed its prediction accuracy and scalability across the two datasets utilized in the case studies (Section~\ref{sec:06_Evaluation_CaseStudy}). 
\wyf{
Employing the prediction pipeline described earlier, we focused on the top $K$ topics in the predictions for each type of impact, ensuring these topics collectively cover more than 80\% grants in each prediction task. We then filtered out topics with fewer than $100$ positive samples.
}

\begin{itemize}[leftmargin=10pt,topsep=1pt,itemsep=0px]
    \item \textbf{AUC}. We present the AUC results on the test sets for the two case studies across different types of impacts in Fig.~\ref{fig:case1_case2_auc}. The figure shows that overall prediction performance is good and remains robust for all types of impacts. 
    \item \textbf{Scalability}. XGBoost is known for its scalability and efficiency~\cite{chen2016xgboost}. 
    We also ran the codes in parallel for different topics to accelerate the prediction process. Moreover, the SciBERT embeddings and prediction model are pre-executed and do not impact the visualization system in real time.
\end{itemize}

\subsection{Case Study} 
\label{sec:06_Evaluation_CaseStudy}
\wyf{We invited our experts to explore the system and then summarized their observations and formed two case studies to demonstrate our system.} 

\subsubsection{\wyf{Funding Impact for a Funding Agency}} 
\label{sec:06_Evaluation_CaseStudy_Case1_NIH} 
We first demonstrated our system from a funder's perspective, engaging our experts \ea and \eb and showcasing the ability of \systemname to help them access the funding impact and inform investment decisions. 

\textbf{Overview of the funding portfolio.} 
The experts first query grants from a federal funding agency (\textbf{\tasktypeoneshort1}, Fig.~\ref{fig:system-ui}(A)) and start from the \grantview and \piview to get an overview of all funded projects. 
This agency has a focus on four main fields~\wyf{(\textbf{\tasktypeoneshort2}, Fig.~\ref{fig:system-ui}(B))}.
Zooming in on these grant fields, \ea noticed a gender disparity across some of the topics they fund, where female PIs appear underrepresented~\wyf{(\textbf{\tasktypeoneshort3}, Fig.~\ref{fig:system-ui}(C1))}. 
Looking into the radar chart for details, he found that projects by female PIs were on par and, in some cases surpassed those by male PIs in certain impact measures (\eg, policy impact). 
This insight about gender disparity clearly caught the attention of the expert, \textit{``We should definitely look more into this.''}

\textbf{Funding impact.} 
The expert \ea was very excited to explore the societal impacts of their grants, as these impact measures represent entirely new information for him. 
Indeed, given the novelty of the data and the linkages, the various impact dimensions offered by \systemname present information that has not been available to decision-makers like \ea, who described himself as ``blind'' to these outcomes. 
\ea turned to \impactselectionview~\wyf{(\textbf{\tasktypetwoshort1}, Fig.~\ref{fig:system-ui}(E))}, selecting the broad impact. 
He was thrilled to learn that the marketplace, clinical, and policy impacts of their grants were notably higher than the global baseline. 
For further details about these types of impacts, he went to the \mainview~\wyf{(\textbf{\tasktypetwoshort1}, Fig.~\ref{fig:system-ui}(D))}. 
The \impactglyph allowed \ea to focus on the field with the largest proportion of funding expenditures which has high impacts of the three impact types~\wyf{(Fig.~\ref{fig:system-ui}(D1))} and to zoom in on the detailed topic-level impacts. 
The system revealed that the funded research had been cited in many policy topics (\eg, ``Virus Disease'', Fig.~\ref{fig:system-ui}(D2)). Although the agency mainly funds research in the US, he was delighted to learn that the policy impact of their funded projects extends across the globe, with numerous policy citations from the Germany, United Kingdom, and other intergovernmental organizations (IGOs)~\wyf{(\textbf{\tasktypetwoshort2}, Fig.~\ref{fig:system-ui}(F1))}, so as the patent impact~\wyf{(Fig.~\ref{fig:system-ui}(F2))}. He exclaimed,
\textit{``My boss should see this!''} 
Our interactions with \ea suggest that \systemname provides him with hard data that can substantially inform their decision-making. We were especially delighted to see that some of the insights were derived from novel data linkages that offered relevant yet previously unknown information to key decision-makers in science, like \ea.

\begin{figure} 
 \centering 
 \includegraphics[width=0.8\linewidth]{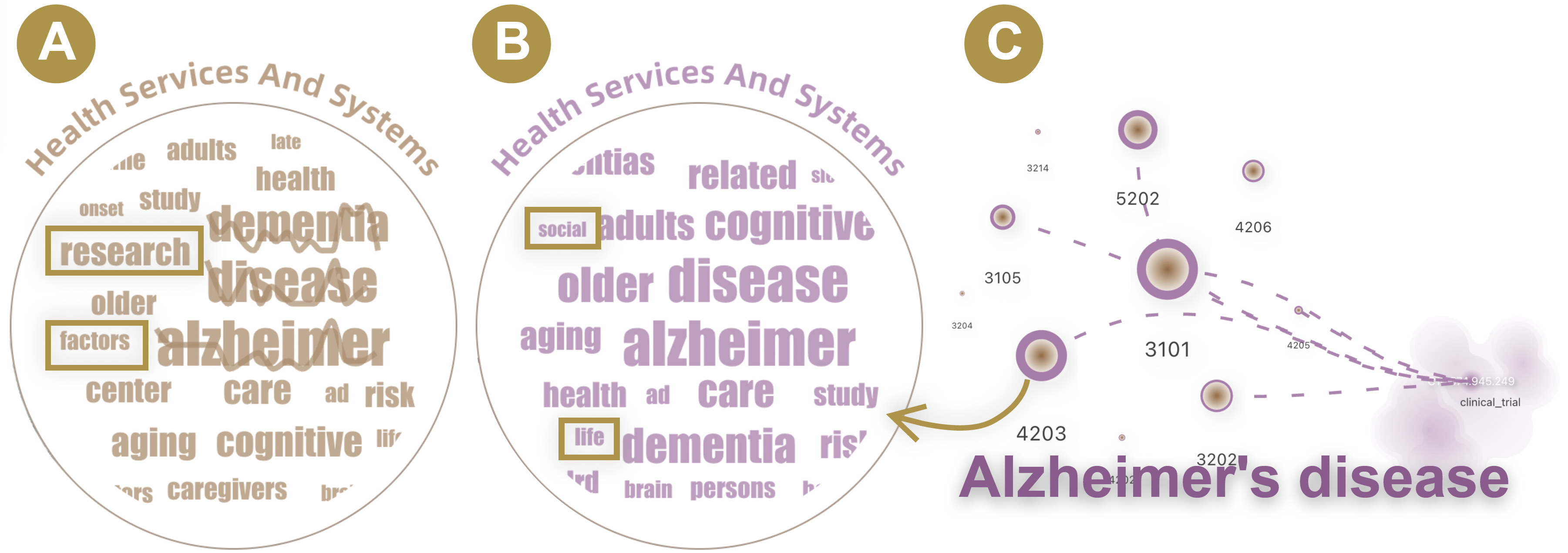}
 \caption{\wyf{Predicted result in Case 1. 
  (A) Historically, grants with impacts on Alzheimer's clinical trials have focused on disease understanding. (B) The recent grants with high prediction scores are oriented toward social and life dimensions of the disease. (C) The topic distribution of recent high-prediction score grants (purple rings) to have an impact on Alzheimer's disease clinical trials. 
 }
 } 
 \label{fig:case1}
\end{figure}


\textbf{Future investment guidance.} 
Whereas \ea is from a public funder, our expert \eb comes from a private investment firm that specializes in investing in biomedical research in academia to advance healthcare. 
\eb was particularly interested in Alzheimer's research, probably due to the increased investment in this area by his organization. 
Here he was drawn to our predictive analytics, which showcases the predicted clinical impacts of current projects related to Alzheimer's disease~\wyf{(\textbf{\tasktypethreeshort1}, Fig.~\ref{fig:case1}(C))}. 
Insights from predictive analytics are, in general, new to decision makers and must be interpreted with care. 
Nevertheless, by looking at the detailed grant keywords, \eb noticed an important trend. 
Whereas the research on Alzheimer's disease that has high clinical impacts over the past twenty years focused primarily on disease understanding~\wyf{(Fig.~\ref{fig:case1}(A))}, 
the predictive analytics suggest that current grants with high predicted clinical impacts (i.e., high prediction scores) seem to be more concentrated on the social and life dimensions of the disease~\wyf{(Fig.~\ref{fig:case1}(B))}. 
Musing over this trend, \eb noted, \textit{``This is very interesting. I wonder if we should look more into combining medical treatment with social support systems.''}
\eb also consulted the \piview~\wyf{(Fig.~\ref{fig:system-ui}(C2))} and noticed several top PIs with relevant research who had not yet been funded by his organization. 
The system allowed him to easily navigate to these PIs' research profiles to assess potential fit and collaboration opportunities. 
\todiscuss{
Here, Alzheimer’s disease is presented merely as an illustrative example to demonstrate the system’s predictive capabilities, and similar analyses could be conducted for other diseases or research areas.
}

Overall, our system offers domain experts key information that they were previously unaware of. This knowledge broadens and sharpens the search space over which they evaluate investments (in ideas or people). While funding or investment decisions depend on a large variety of factors and considerations, \systemname helps decision makers reduce key blind spots and ultimately offers a better and more comprehensive set of data to inform evidence-based decision making.

\subsubsection{\wyf{Funding Landscape for the Visualization Community}} 
\label{sec:06_Evaluation_CaseStudy_Case2_VIS} 
Given the growing interest in interdisciplinary research~\cite{park2023interdisciplinary}, our SciSci experts (\ec-\ee) examined visualization field grants to illustrate the diverse impact of global funding on interdisciplinary research. 

\begin{figure*} [!htb]
 \centering 
 \includegraphics[width=1\linewidth]{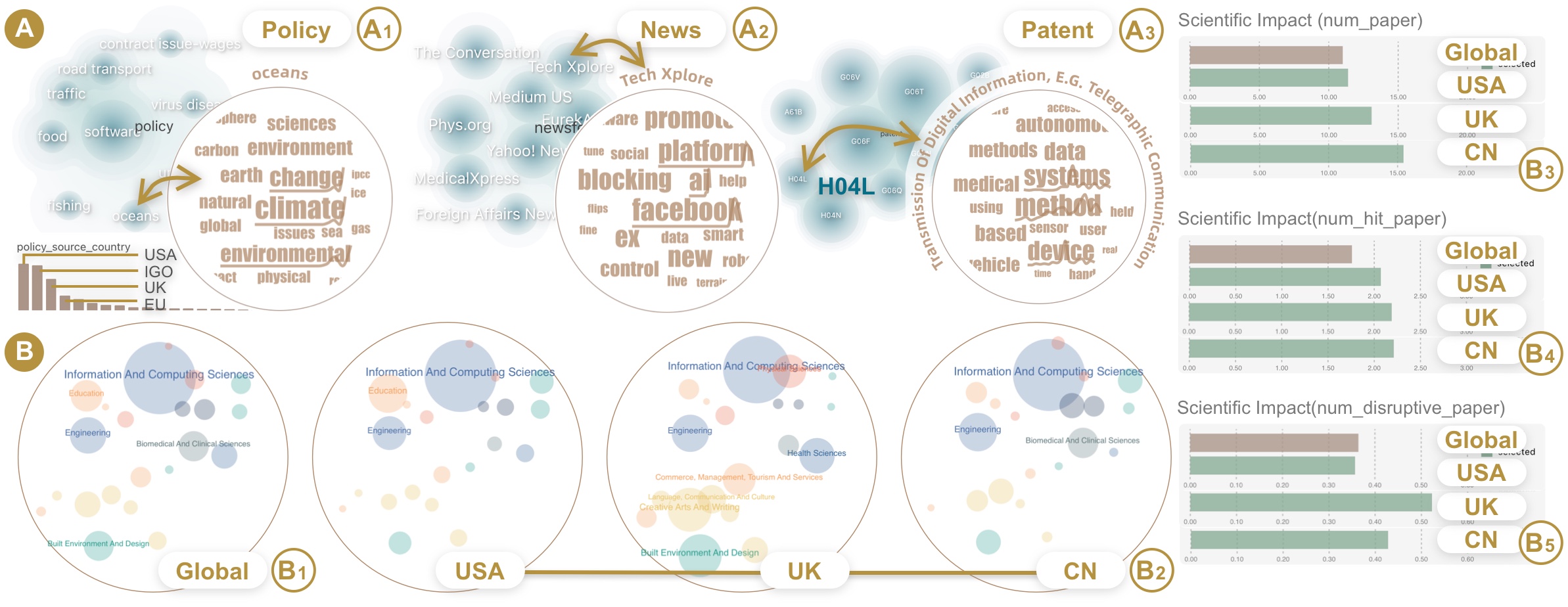}
 \caption{
    The impacts of funding for Case 2. 
    (A) The broader impacts of funded visualization research include the policy impact for climate change (A1), media impact for AI (A2), and marketplace impact for digital communication (A3). 
    (B) The comparison of funded visualization research in different countries shows interesting variation between the fields that are funded ((B1)-(B2)) and the science impact ((B3)-(B5)). 
 } 
 \label{fig:case2}
\end{figure*}

\textbf{Overview of the funding impact. } 
The experts began their exploration of the system by looking at an overview of the funded projects~\wyf{(\textbf{\tasktypeoneshort2}, Fig.~\ref{fig:case2}(B1))}. 
Reflecting the highly interdisciplinary nature of the vis community, the funded projects cover a wide range of topics beyond computer science, such as education, biomedicine, and design. 
Exploring the broad impact of these projects, they found that the projects have had a significant influence on transportation, environmental, and health policies~\wyf{(\textbf{\tasktypetwoshort1}, Fig.~\ref{fig:case2}(A1))}. 
Digging deeper into the policies about oceans, they discovered that these policies had a strong emphasis on climate change and that this impact was seen in the policy documents of multiple countries, as well as IGOs ~\wyf{(\textbf{\tasktypetwoshort2}, Fig.~\ref{fig:case2}(A1))}. An expert noted that the system documents a contribution of visualization research that had not necessarily been recognized in the past: 
\textit{``Interesting... This suggests that vis research plays a role in addressing key societal challenges such as climate change, which is currently underappreciated.''}
%
The experts then turned to the media impact~\wyf{(\textbf{\tasktypetwoshort1}, Fig.~\ref{fig:case2}(A2))}. 
Both specialized and general-audience media organizations have covered vis research, touching on a wide range of topics, including AI and pandemics. The breadth of coverage reflects the far-reaching media interest in and the applicability of vis research. While experts may have been aware of this broad appeal at a high level, \textit{FtF} provided them with concrete evidence of this range and specificity that they had not seen before. 
The system also allowed the experts to compare the direct and indirect outcomes of grants in terms of patents~\wyf{(\textbf{\tasktypetwoshort1}, Fig.~\ref{fig:case2}(A3))}. 
They discovered a new patent category in the top frequent patent fields in indirect impact: 
transmission of digital information (H04L).
\textit{``It seems that researchers in vis don't typically file patents in this technology topic, but their scientific expertise is widely leveraged in related areas,''} noted an expert, seeing an \textit{``opportunity in leveraging vis for digital communication technologies.''}

\textbf{National disparities in visualization research.} 
The experts were also interested in exploring national differences in funded vis projects and their impacts. 
For this, the experts compared three countries: the United States, the United Kingdom, and China. 
Most funding for vis projects goes to the computer science topic, but there were notable differences between the countries in the distribution of other fields~\wyf{(\textbf{\tasktypeoneshort2}, Fig.~\ref{fig:case2}(B2))}: the U.S. prioritizes education, the U.K. favored art and design, and China emphasized biomedical fields. 
Turning to the scientific impacts, though vis projects in the three countries all tend to be greater than the global average, the experts found that there were notable variations across countries that they had not seen before~\wyf{(\textbf{\tasktypetwoshort1}, Fig.~\ref{fig:case2}(B3)-(B5))}. 
The U.K. showed the highest propensity for disruptive science among the three countries~\wyf{(Fig.~\ref{fig:case2}(B5))}. \wyf{Interestingly, vis papers from China feature a higher fraction of disruptive work than the U.S.. When the experts combined this insight about disruptive science with other related information the system offered, including the hit papers statistics, they learned that vis funding from China tends to yield more papers per grant (Fig.~\ref{fig:case2}(B3)), but not substantially more hit papers than most other countries (Fig.~\ref{fig:case2}(B4)). 
}

In this case, \textit{FtF} offers experts a wide range of insights---from the interdisciplinarity of vis work to national differences in the impact.

\subsection{Expert Interview}
\label{sec:06_Evaluation_ExpertInterview} 
In addition to feedback from the experts described in Section~\ref{sec:03_Background_AnalysisTask}, we presented the system to new experts, including a founding director of a leading university's Technology Transfer Office (\newea) and senior researchers who focus on the impact of science on public policy and health (\neweb and \newec).
We also interviewed three visualization experts to potentially identify other uses of the system beyond its applications for funding agency decision-making, including a senior Ph.D. experienced in grant writing (\newed) and two graduates with over three years of research experience (\newee and \newef). 
Each interview with these new experts lasted 90 minutes, covering project introductions, system and visual design illustrations (using case 1), free time to explore the system, and a semi-structured interview. Experts were encouraged to share their thoughts and impressions during the process. We summarize the feedback from the two groups of experts below. 

\textbf{Analysis Workflow.} 
All experts reported that the system's workflow was straightforward and that they could complete the analysis tasks, from data query and funded project analysis to the evaluation of impact. 
\eb praised the system's sophistication and its advances beyond the capabilities offered by their daily used tools like WOS and Scopus. 
\ea was also excited about the system's ability to identify research topics and PIs using prediction results: \textit{``This changes the game. Instead of passively waiting for proposals, I can use the tool as a radar to scan all suitable topics and PIs to invest in.''} 

\textbf{Investment prediction and impact metrics.} 
\ea appreciated the prediction feature that allowed him to effectively identify potential investment opportunities, and he commended the impact metrics for their clarity and usefulness. 
\ea also praised the relative impact index and the  \impactglyph, which he saw as valuable tools for quickly identifying the strengths and weaknesses of a group of grants. However, he called for greater transparency in the prediction model.
Voicing similar enthusiasm, 
\newec appreciated the PI and grant field level impact measure to compare high-level funding impacts such as funders and countries. 

\textbf{Visualization.} 
The experts agreed that the visualization components satisfy all of the analysis tasks. 
\eb, \newea, and \newee were especially impressed with the citation flows in the \mainview that features an intuitive presentation of the impact of funding at the topic level. 
\newed commented that the word cloud design was both readable and easy to understand. 
Many of the experts appreciated the \impactglyph. One observed, \textit{``the ripple metaphor shows the multi-dimensional impact in a vivid way.''}
Nevertheless, \ea and \newea noted that there would be a learning curve in their use of the glyph, acknowledging,  
\textit{``it's a bit new to me.''} All experts were able to understand and use it effectively, however, after being given some time to explore the system. One expert affirmed, \textit{``I think it will be very useful to compare group level grants, as the difference among glyph shapes is quite apparent.''}

\textbf{Suggestions.} 
The experts saw the current metrics as straightforward and useful for analyzing the impact of funding, but they had some thoughtful suggestions adding more innovative measures such as ``risk.'' 
\ea also recommended adding a raw data table that presents both grants and impact documents for deeper analysis. 
\eb pointed out potential limitations in the data, \textit{``many patents do not directly indicate the support of grants. We need to do deep data cleaning to search the whole patent doc to see whether they have claimed grant support.''}

\section{Discussion}
\label{sec:07_Discussion} 
Our study illustrates how connecting funding to its downstream impacts opens new possibilities for both research and practice. By linking grants to publications, patents, clinical trials, policy, and media at scale, \systemnamelong (\systemname) provides not only a descriptive map but also an integrative infrastructure for future work in the science of science and science policy.

The availability of large-scale data linking upstream funding to diverse downstream impacts presents new research opportunities. With \systemname, researchers can examine how different funding models shape the trajectory of science and innovation, compare funding portfolios across nations, and study the efficiency and time lags of investment returns across domains. The system can also support investigations into long-standing questions. For example, how public versus private funders differ in their societal footprint, or whether funding strategies that emphasize interdisciplinarity systematically yield broader impacts. In this sense, \systemname represents not an end point but a foundation for new lines of work across the science of science, innovation studies, and policy research.

Equally important, \systemname was built to be actionable by non-technical stakeholders. Policymakers, program officers, university leaders, and even philanthropic funders face growing pressure to justify investments, prioritize areas of opportunity, and demonstrate societal relevance. By making complex data accessible and interpretable, \systemname equips these decision-makers to explore their portfolios, identify blind spots, and consider new directions. For example, program officers might use the system to surface emerging investigators with outsized potential; university leaders might assess how their institution’s grants contribute to policy or market outcomes; and policymakers might gain perspective on whether federal portfolios align with national and regional priorities. The diversity of these use cases underscores the broad value of making SciSci insights transparent and navigable.

While our case studies focus on selected funders and domains, the underlying datasets and framework are global in scope. The system is not limited to U.S. agencies but can, in principle, extend to philanthropic organizations, multinational funding programs, and national science foundations worldwide. This breadth makes \systemname a potential global public good: a resource for comparing funding impacts across borders and for enabling international comparisons at a moment of intensifying global competition in science and technology.

As a prototype, \systemname has several limitations. Our predictive models, while effective, can be further improved with more data and more advanced machine learning methods, and can benefit from further development to maximize transparency and interpretability, particularly for stakeholders unfamiliar with the technical aspects. 
In addition, while our datasets are among the most comprehensive, they are inevitably shaped by coverage biases, such as the predominance of English-language sources in policy documents and newsfeeds, which may underrepresent impacts in other linguistic and regional contexts. 
Further, while our evaluation demonstrates the system’s usability, the evaluation should be considered as exploratory, and future work may further expand the datasets, integrate additional impact measures, and tailor customized visual workflows to distinct user groups, from administrators and leaders to data scientists and researchers. Addressing these challenges will improve the robustness of the system, further enhancing its value as an infrastructure for both decision-making and research.

Taken together, \systemname demonstrates how the combination of SciSci insights and visualization design can help address a long-standing gap between funding inputs and societal outcomes. At a time when science budgets face scrutiny, geopolitical competition intensifies, and public trust in science is contested, the ability to assess and communicate the multidimensional impacts of funding is critical. By providing a transparent, scalable, and extensible framework, \systemname offers both a practical resource for decision-makers and a foundation for future research into the broad impacts of science and science investment.

\section*{Data Availability}
\label{sec:09_DataAvailability} 
Data necessary to reproduce all plots will be made freely available.

\section*{Code Availability}
\label{sec:09_CodeAvailability} 
Code necessary to reproduce all plots will be made freely available.
\end{spacing}

\newpage

\bibliographystyle{naturemag}
\bibliography{main}

\acknowledgments{%
We thank all members of the Center for Science of Science and Innovation (CSSI) at Northwestern University for helpful discussions, and Alyse Freilich for editing. This work is supported by the National Science Foundation (award number 2404035) and the Future Wanxiang Foundation. Any opinions, findings, and conclusions or recommendations expressed in this material are those of the author(s) and do not necessarily reflect the views of the National Science Foundation and the Future Wanxiang Foundation. 
}










\end{document}